\documentclass[10pt,aps,showpacs,a4paper,nofootinbib]{revtex4}

\usepackage{graphicx}
\usepackage{graphics}

\newcommand{\half}{\textstyle{\frac{1}{2}}}

\begin{document}

\title{Azimuthal asymmetries in unpolarized Drell-Yan events on nuclear
targets}

\author{Andrea Bianconi}
\email{bianconi@bs.infn.it}
\affiliation{Dipartimento di Chimica e Fisica per l'Ingegneria e per i Materiali, 
Universit\`a di Brescia, and \\
Istituto Nazionale di Fisica Nucleare, Sezione di Pavia, I27100 Pavia, Italy}

\author{Marco Radici}
\email{radici@pv.infn.it}
\affiliation{Dipartimento di Fisica Nucleare e Teorica, Universit\`{a} di Pavia, 
and\\
Istituto Nazionale di Fisica Nucleare, Sezione di Pavia, I-27100 Pavia, Italy}

\begin{abstract}
We show that for Drell-Yan events by unpolarized hadronic projectiles and nuclear
targets, azimuthal asymmetries can arise from the nuclear distortion of the hadronic
projectile wave function, typically a spin-orbit effect occurring on the nuclear
surface. The asymmetry depends on quantities that enter also the spin asymmetry in
the corresponding Drell-Yan event on polarized free nucleonic targets. Hence, this
study can be of help in exploring the spin structure of the nucleon, in particular
the transverse spin distribution of partons inside the proton. All arguments can be
extended also to antinucleon projectiles and, consequently, apply to possible future
measurements involving nuclear targets at the foreseen HESR ring at GSI.
\end{abstract} 

\pacs{13.85.Qk,13.88.+e,24.10.Ht,25.43.+t}

\maketitle

\section{Introduction}
\label{sec:intro}

High-energy collisions of hadrons can be a very interesting source of information for 
testing QCD in the nonperturbative regime. In fact, in the past years several 
experiments have been performed that still await for a correct interpretation of the
resulting data. Among others, hadronic collisions on transversely polarized proton 
targets~\cite{Bunce:1976yb,Adams:1991cs,Bravar:1999rq,Adams:2003fx}, where an 
azimuthally asymmetric distribution of final-state products (with respect to the normal
of the production plane) is observed when flipping the transverse spin of the target or 
of the final products, the so-called transverse spin asymmetry. Perturbative QCD cannot 
accommodate for such asymmetries, sometimes as large as 40\% also at high 
energy~\cite{Bunce:1976yb}. This problem has triggered a growing interest in this field 
of hadronic spin physics; new data and rapid developments are foreseen in the near 
future, as it emerged in a recently devoted workshop~\cite{cerncourier} (for a review 
covering also processes with lepton beams, see 
Refs.~\cite{Jaffe:1996zw,Barone:2003fy}). 

\begin{figure}[h]
\centering
\includegraphics[width=7cm]{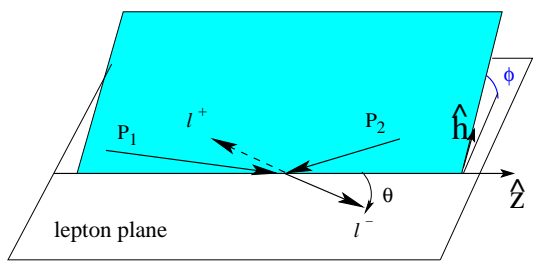}
\caption{The Collins-Soper frame.}
\label{fig:dyframe}
\end{figure}

Here, we will concentrate on a series of measurements involving nuclear targets,
namely high-energy collisions of pions and antiprotons on various unpolarized 
nuclei~\cite{Falciano:1986wk,Guanziroli:1987rp,Conway:1989fs,Anassontzis:1987hk} where the 
cross section for Drell-Yan events shows an unexpected largely asymmetric azimuthal 
distribution of the final lepton pair with respect to the production plane. It is usually 
convenient to study the problem in the so-called Collins-Soper frame~\cite{Collins:1977iv} 
of Fig.~\ref{fig:dyframe}. A lepton plane is defined by the direction of emission of the 
lepton pair and by the $\hat{z}$ axis which is approximately formed by the average of the 
directions of the colliding hadron momenta ${\bf P}_1$ and ${\bf P}_2$ (for a more 
rigorous definition, see Ref.~\cite{Boer:1999mm}). Azimuthal angles are measured in a 
plane perpendicular to $\hat{z}$ and containing the axis $\hat{\bf h} = 
{\bf q}_{_T}/|{\bf q}_{_T}|$, where ${\bf q}_{_T}$ is the transverse momentum of the final
lepton pair detected in the solid angle $(\theta, \phi)$. 

At leading order in $\alpha_s$, the cross 
section~\cite{Falciano:1986wk,Guanziroli:1987rp,Conway:1989fs,Anassontzis:1987hk} can be 
parametrized as
\begin{equation}
\frac{1}{\sigma}\,\frac{d\sigma}{d\Omega} = \frac{3}{4\pi}\, \frac{1}{\lambda + 3}\, 
\left( 1 + \lambda \, \cos^2\theta + \mu \, \sin^2\theta \, \cos \phi + \frac{\nu}{2}
\, \sin^2\theta \, \cos 2\phi \right) \; , 
\label{eq:unpoldyx}
\end{equation}
with $\lambda \sim 1$ and $\mu \ll \nu \sim 30$ \%. Both Leading Order (LO) and 
Next-to-Leading Order (NLO) perturbative QCD calculations give $\lambda \sim 1$ and  
$\mu \sim \nu \sim 0$~\cite{Brandenburg:1993cj}, because they are based on the 
assumption of massless quarks and collinear annihilation. In fact, this result holds 
only if the annihilation direction is well defined and coincides with the $\hat{z}$ axis 
of Fig.~\ref{fig:dyframe}, because the cross section would depend only on $\theta$. In 
general, the assumption of massless quarks leads to the so-called Lam-Tung sum rule 
$1-\lambda = 2\nu$~\cite{Lam:1980uc}, which is badly violated by data. More complicated 
mechanisms, like higher twists or factorization breaking terms at NLO, are not able to 
explain the size of $\nu$ in a consistent 
picture~\cite{Brandenburg:1994wf,Eskola:1994py,Berger:1979du}. A promising 
interpretation relies on the observation that at leading twist the cross
section for unpolarized Drell-Yan, when kept differential also in $d{\bf q}_{_T}$, 
contains a term proportional to $\cos 2\phi \, ( h_1^\perp \, \otimes \, 
\bar{h}_1^\perp)$, i.e. a specific convolution of the distribution function $h_1^\perp$ 
that describes the transverse momentum distribution of transversely polarized partons 
inside unpolarized hadrons~\cite{Boer:1999mm}, while $\bar{h}_1^\perp$ describes the 
annihilating antiparton partner. 

However, it is important to stress that the above results are obtained using nuclear 
target: before getting to the collision with a bound nucleon that generates the 
Drell-Yan event, the hadronic beam crosses the nuclear surface and travels inside the 
nuclear medium for some length depending on its energy. The aim of this paper is to try 
to understand whether effects due to the beam-nucleus interaction may assume a relevant 
role in contributing to the observed azimuthal $\cos 2\phi$ asymmetry. In other words, 
if an intrinsic transverse momentum distribution can affect the elementary annihilation 
of partons leading to an asymmetric azimuthal distribution of final lepton pairs 
through the function $h_1^\perp$, then a hadron deflected by nontrivial Initial-State 
Interactions (ISI) with the target nucleus can transfer its transverse momentum to the 
final lepton pair and contribute to the same azimuthal asymmetry. More, we will see 
that such ISI can originate azimuthal asymmetries also when no asymmetry is produced at 
the level of a free proton target.

The highly energetic pions and antiprotons used in measurements reported in 
Refs.~\cite{Falciano:1986wk,Guanziroli:1987rp,Conway:1989fs,Anassontzis:1987hk} demand a 
description of their internal structure in terms of elementary degrees of freedom (quarks 
and gluons) in order to correctly describe their propagation and to study effects that can 
modify high-energy azimuthal asymmetries at RHIC or LHC (see, for example, 
Refs.~\cite{Fries:2000da,Kopeliovich:2001hf}). Here, we will consider Drell-Yan processes 
at lower energies of interest for the foreseen HESR ring at GSI, where collisions of 
(polarized) protons and antiprotons with energy in the range $10 \div 30$ GeV will be
studied~\cite{panda,pax,pax2,assia,Efremov:2004qs,Anselmino:2004ki,Radici:2004ij}. Hence, 
we will adopt a description based on "traditional" degrees of freedom as nucleons and 
pions; this assumption is justified in the following Sec.~\ref{sec:eik}.

The paper is organized as follows. In Sec.~\ref{sec:eik}, the nuclear damping of the
projectile wave function, based on the eikonal approximation, is reviewed and the 
induced modifications to the scattering amplitude are discussed. In Sec.~\ref{sec:so}, 
the spin-orbit effect on the nuclear surface is discussed and the formulae are 
implemented with an additional damping that depends on the polarization component of 
the incoming hadron beam. In Sec.~\ref{sec:asym}, it is shown how all this can produce 
azimuthal asymmetries in Drell-Yan events from unpolarized hadron-nucleus collisions 
even when no asymmetry is generated using a free proton target. Finally, in 
Sec.~\ref{sec:size} the discussion is presented about the issues of the size of this 
nuclear effect and of the possibility of disentangling it from genuine asymmetries 
related to the Drell-Yan hard event. Results are summarized in Sec.~\ref{sec:end}.

\section{Spin-independent nuclear damping}
\label{sec:eik}

As already stressed in Sec.~\ref{sec:intro}, we will consider Drell-Yan processes 
involving nuclear targets at energies of interest for the foreseen HESR ring at GSI. At 
present, the discussion is still open about the two options of an antiproton beam of 
energy 15 GeV hitting a fixed proton target or colliding on a proton beam of energy around 
3 GeV, such as the center-of-mass energy ranges from 5 to approximately 14 
GeV~\cite{panda,pax,pax2,assia,Efremov:2004qs,Anselmino:2004ki,Radici:2004ij}. At 
this scale, the use of traditional "nuclear" degrees of freedom (nucleons and pions) is 
still appropriate for describing the propagation of a hadron inside the nucleus, also 
because spin-orbit effects are still significant. Moreover, a distance in space (the 
so-called formation length) must separate the Drell-Yan hard event from the ISI 
rescatterings in such a way that the two processes do not influence each other; the 
formation length increases with energy but it is already significant at 15 GeV beams. 
In fact, the same problem occurs in the semi-inclusive quasi-elastic electroproduction 
of protons on nuclei about the Final-State Interactions (FSI) of the proton travelling 
inside the residual nuclear medium after the hard event of virtual photon absorption. 
The NE18 experiment~\cite{NE18-1} about the $^{12}$C$(e,e'p)X$ reaction has shown that, 
at 10 GeV energies, the ejectiles interact with the nuclear medium exactly as "normal" 
protons, before emerging and being actually detected as protons. 

At the same time, a projectile with energy 15 GeV impinging on a
target nucleus is energetic enough such that the collision can be safely considered in 
the Born Approximation. The main effect of the nuclear medium is, therefore, to distort 
the projectile wave function. The most popular and widely adopted approach to such a 
situation is the Glauber method~\cite{glau59}, which has a long well established 
tradition of successfull results in the field of high-energy proton-nucleus elastic 
scattering (for a review, see also Ref.~\cite{pprev78}). Again, a parallel with 
quasi-elastic $A(e,e'p)X$ reactions can be established, since above the inelastic 
production threshold, i.e. for proton momenta $p \gtrsim 1$ GeV/$c$, it has been 
shown~\cite{br95,br96a,br96b} that this method gives the same FSI effects as the usual 
approach based on the Distorted Wave Impulse Approximation (DWIA), where the projectile 
wave function is expanded in partial waves and a Schr\"odinger equation is solved for 
each partial wave $L$ with a complex optical potential up to a $L_{\rm max}$ depending on 
the projectile energy and satisfying a stringent convergency criterion (for a review, see 
Ref.~\cite{bgprbook}). Actually, it was directly checked that the equivalence of these 
methods holds when the Glauber scattering wave is replaced by an eikonal wave 
function~\cite{br97}. Because of the similarity between the FSI problem in $A(e,e'p)$ 
reactions and the ISI problem in $p-A$ collisions, we will adopt here the same eikonal 
approximation for the projectile wave function.

Since for a fastly moving object the nuclear density can be considered roughly constant 
but for a small portion corresponding to the nuclear surface, the eikonal wave function 
of the projectile with momentum ${\bf p}$ can be represented by the damped plane wave
\begin{equation}
\psi ({\bf r}) \approx e^{-{\bf p}_I\cdot {\bf R}} \, e^{i {\bf p}\cdot {\bf r}} \, 
e^{-{\bf p}_I\cdot {\bf r}} \; ,
\label{eq:eikwf}
\end{equation}
which corresponds to an approximate solution of the Schr\"odinger equation inside 
homogeneous nuclear matter with $|{\bf p}_I|=p_I$ proportional to the ratio between the 
absorptive (imaginary) potential and $|{\bf p}|=p$; it is equivalent to neglect the
solution of the Schr\"odinger equation that propagates backward in space~\cite{glau59}. 
For a non-homogeneous nucleus, the same approximation leads to the Glauber wave function. 
The factor $\exp (-{\bf p}_I\cdot {\bf R})$ is due to the incoming proton wave being 
properly normalized to 1, where ${\bf R}$ is a constant vector whose modulus gives the 
nuclear radius. A reasonable value for $p_I$ can be estimated by reproducing, among other 
things, the NE18 data~\cite{NE18-1} for the ${}^{12}$C$(e,e'p)$ reaction where the flux of 
the outgoing proton with $p = 1.4$ GeV/$c$ is overall reduced by 40\%, including the $Q^2$ 
behaviour of the related transparency coefficient~\cite{gol98}. It turns out that 
$p_I \ll p$, for the above NE18 kinematics typically $p_I \sim 50$ MeV/$c$. Nevertheless, 
this damping is able to generate "geometric" FSI responsible for the asymmetric proton 
knockout with respect to the scattering plane in exclusive ${}^{12}$C$(\vec{e},e'p)$ 
reactions with a polarized lepton beam (the so-called fifth structure function 
problem)~\cite{br97}.

The eikonal approximation of Eq.~(\ref{eq:eikwf}) is, by construction, a high-energy
approximation and its reliability increases with the projectile energy, ideally
corresponding to the limit of an infinite number of partial waves for the solution in
the DWIA. In these conditions, as a first step we neglect spin constributions to
Eq.~(\ref{eq:eikwf}) (but we will reconsider the problem later). Moreover, if the
projectile propagates along the $\hat{z}$ direction, we will assume that also the
damping is uniformly concentrated along the same direction, i.e. ${\bf p}_I
\parallel {\bf p} = p \hat{z}$; consequently, 
\begin{equation}
\psi ({\bf r}) \approx e^{-p_I R} \, e^{i p z} \, 
e^{-p_I z} \equiv \psi (z) \; . 
\label{eq:eikwfz}
\end{equation}
This introduces a (small) error in the asymptotic angular distribution of scattered
projectiles, but to our purpose of considering only hard events inside the target
nucleus it is irrelevant. 

The function~(\ref{eq:eikwfz}) can be considered as a plane wave with the complex
momentum ${\bf p} + i {\bf p}_I = (p + i p_I) \, \hat{z} = P \, \hat{z} = 
{\bf P}$~\cite{br97}. Therefore, as a simple plane wave gets Fourier transformed to a 
$\delta$ distribution, similarly the "complex" plane wave $\exp{(i P z)}$ can be Fourier 
transformed by extending the definition of the $\delta$ distribution to the complex plane. 
Following Ref.~\cite{davy}, since the $\delta$ distribution of a real argument can be 
defined as
\begin{equation}
\delta (x-\bar{x}) = \frac{1}{\pi} \, \lim_{\epsilon \rightarrow 0}\,
\frac{\epsilon}{(x-\bar{x})^2+\epsilon^2} = \lim_{\epsilon \rightarrow 0}\, 
\frac{1}{2\pi i} \, \left( \frac{1}{x-\bar{x}-i\epsilon} - 
\frac{1}{x-\bar{x}+i\epsilon} \right) \; ,
\label{eq:rdelta}
\end{equation}
the analytic continuation to a complex argument becomes
\begin{equation}
\delta (z-\bar{z}) \equiv \delta (z-(\bar{x}+i\bar{y})) = \frac{1}{2\pi i}\, \left(  
\frac{1}{z-\bar{x}-i\bar{y}} - \frac{1}{z-\bar{x}+i\bar{y}} \right)  = 
\frac{1}{2\pi i}\, \left(  \frac{1}{z-\bar{z}} - \frac{1}{z-\bar{z}^*} \right)  \; .
\label{eq:cdelta}
\end{equation}
In other words, when a function $f(x)$ of the real argument $x$ is convoluted with
$\delta (x-\bar{x})$, the integrand is pinched between the two singularities
$x=\bar{x} \pm i\epsilon$, which eventually collapse to the same value in the limit
$\epsilon \rightarrow 0$. For a function $f(z)$ of a complex argument $z$, the effect
is similar if $f(z)$ is analytic and $f(z)\rightarrow 0$ for $|z|\rightarrow \infty$.
In fact, in this case we have
\begin{equation}
\int_C dz \, \delta (z - \bar{z})\, f(z) = f(\bar{z}) \; ,
\label{eq:intcdelta}
\end{equation}
where $C$ is an integration contour in the complex plane extending to Re$(z) \rightarrow
\pm \infty$ on the real axis and closing in the upper plane for Im$(\bar{z}) =
\bar{y} > 0$ (and viceversa). 

How to interpret the $\delta$ distribution with a complex argument? With no damping, the
eikonal wave in momentum space is represented by $\delta (k -p)$, a singular distribution
in $k$ peaked on the momentum $p$ of the projectile. With a nuclear damping $(p_I \neq 0)$,
the momentum distribution is still peaked around $p$ but is no longer singular, since the
potential inside the nuclear volume distorts the wave function and introduces a width
$\Delta p \sim 1 / R$ around $p$. For energies above the inelastic proton-nucleon
threshold, the nuclear potential is mainly imaginary and generates an absorption of the
projectile flux (here represented by the damping $p_I$); hence, $\Delta p \sim 
1/\tilde{z}$, where $\tilde{z}$ is the absorption length if the projectile is propagating 
along the $\hat{z}$ axis. If $\tilde{z} < R$, then $\Delta p$ is mostly given by 
the damping, i.e. by $p_I$. In conclusion, the distribution $\delta (k - p -p_I)$ is peaked
around $p$ but has a finite width $\Delta p$, or equivalently it describes a projectile
with a finite absorption length $\tilde{z} \sim 1/\Delta p$, that is mainly given by the
damping $p_I$.

In order to understand the modifications inside the scattering amplitude, it is useful to
consider first the case without damping; for the projectile-nucleus collision we have 
\begin{equation}
a(p) = \int d{\bf q} \, d{\bf k}\, \psi^*_p ({\bf q})\, A({\bf q},{\bf k}) \approx
\int d{\bf q} \, d{\bf k}\, \delta ({\bf q} - p\hat{z})\, A({\bf q},{\bf k})
= \int d{\bf k}\, A(p\hat{z},{\bf k}) \; , 
\label{eq:pwampl}
\end{equation}
where $\psi_p$ represents the incoming wave function in momentum space centered around $p$,
the function $A$ contains all the dynamics of the hard collision, and the
integration upon $d{\bf k}$ covers the portion of phase space complementary to 
the one pertaining the incoming projectile, i.e. it represents the integration upon
momenta of nucleons participating to the collision and of final lepton products. Then, if 
the function $A$ is well behaved and vanishes for aymptotically large momenta, 
Eq.~(\ref{eq:pwampl}) can be generalized to
\begin{equation}
a(p+i p_I) = \int d{\bf Q} \, d{\bf k}\, \delta ({\bf Q} - {\bf P})\, 
A({\bf Q},{\bf k}) = \int d{\bf k}\, A((p+i p_I) \hat{z},{\bf k}) \; .
\label{eq:dwampl}
\end{equation}

\begin{figure}[h]
\centering
\includegraphics[width=5cm]{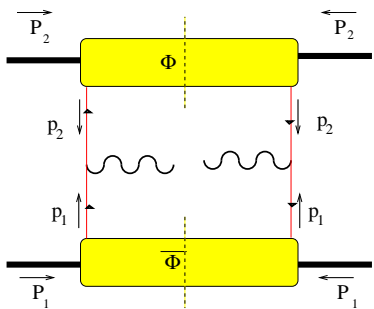}
\caption{The leading-twist contribution to the Drell-Yan process.}
\label{fig:dy0}
\end{figure}

If the invariant mass $M$ of the Drell-Yan lepton pair is kinematically constrained in a 
range where the annihilation proceeds through a virtual photon, the dominant contribution 
at leading twist is represented by the handbag diagram of 
Fig.~\ref{fig:dy0}~\cite{Ralston:1979ys}, where the blobs represent all the possible 
residual states of the two hadrons after the hard event that are not observed. Hence, a 
sum over all possible configurations is understood in the two cuts. As such, the process 
is by construction semi-inclusive and the cross section will be proportional to an 
incoherent sum of several transition probabilities, each one leading to the same final 
lepton pair. Neglecting the irrelevant phase space coefficient, the cross section with no 
nuclear damping will be related to 
\begin{equation}
W^o (p) = \sum_n \, \vert a_n(p)\vert^2 \; ,
\label{eq:xsect}
\end{equation}
where the number of terms depends on the model used to describe the blobs in 
Fig.~\ref{fig:dy0}. 

From Eq.~(\ref{eq:dwampl}), it is evident that the damping of the projectile wave 
function implies a shift in the argument of the elementary scattering amplitude, namely 
$a_n(p) \rightarrow a_n(p+i p_I)$. Since $p_I \ll p$, we have
\begin{equation}
a_n(P=p+i p_I) = a_n(p) + i p_I \frac{\partial a_n}{\partial P} \Big\vert_{p_I=0} + 
o(p_I) \; ,
\label{eq:taylor}
\end{equation}
so that Eq.~(\ref{eq:xsect}) gets modified into
\begin{equation}
W(p,p_I) = W^o (p) - 2 p_I {\rm Im} [G(p)] +o(p_I) \; ,
\label{eq:xsect2}
\end{equation}
with
\begin{equation}
G(p) = \sum_n \, \left( a_n^*\,\frac{\partial a_n}{\partial P}\right) 
\Big\vert_{p_I=0} \; .
\label{eq:fsi}
\end{equation}
Equation~(\ref{eq:xsect2}) immediately suggests that, when building an azimuthal
asymmetry of the type described in Sec.~\ref{sec:intro}, if for some reason the leading 
term $W^o (p)$ gets cancelled, an asymmetry could arise from the projectile-nucleus 
interaction through the interference term $G(p)$. 

Therefore, small contributions $(p_I \ll p)$ can become important in an asymmetry. 
Consequently, we have to go back to the problem of the spin dependence of the 
projectile-nucleus interaction, which was initially neglected.

\section{Spin-orbit effects}
\label{sec:so}

The eikonal approximation is based on the assumption that a fast moving object experiences 
a roughly constant nuclear density but on the nuclear surface, where modifications of the 
projectile wave function can happen due to soft spin-orbit interactions. However, it is
well known that the spin-orbit potential $V_{so}$ is proportional to the gradient of the
nuclear density $\rho$. For a projectile travelling along the usual $\hat{z}$ axis, the
spin-orbit contribution $\psi_{so}$ to its wave function in the eikonal approximation reads
\begin{equation} 
\psi_{so} (z) \approx \int_{-\infty}^z dz' \, V_{so}(z') \propto \int_{-\infty}^z dz' \, 
\frac{d\rho}{dz'} = \rho (z) \; .
\label{eq:sopot}
\end{equation}
In other words, the modifications to the wave function induced by the spin-orbit potential 
do not depend on the details of the surface of the nucleus but, rather, on its bulk 
density. 

Since the projectile momentum is ${\bf p} = p\,\hat{z}$, it is natural to quantize the 
spin along the $\hat{y}$ direction such that the operator form of the spin-orbit potential 
is
\begin{equation}
V_{so}({\bf r}) \, \mbox{\boldmath $\sigma$} \cdot {\bf r} \times {\bf p} \propto 
- \frac{d\rho}{dz} \, p \, x \, \sigma_y \; ,
\label{eq:sooper}
\end{equation} 
where $\mbox{\boldmath $\sigma$}$ is the vector of 2x2 Pauli matrices. Consequently, the 
two spin components $\psi_\pm$ of the incoming wave function (if the projectile has spin 
$\half$) feel a spin-orbit potential of opposite sign; at large energies, $V_{so}$ is also 
purely absorptive so that the net effect is that different polarization components of the 
incoming hadron beam feel different dampings. In the given conditions, the asymmetry 
should be reasonably independent from $\hat{z}$, because it is created when the projectile 
enters the nuclear surface, and it is along the $\hat{x}$ direction: the nucleus is 
divided in two emispheres (e.g., $x>0$ and $x<0$ in the frame where the target is at rest 
and centered in the origin), each one being crossed by a flux of projectiles with opposite 
polarization. 

Since the spin-orbit damping adds to the nuclear damping described in the previous section,
we can formally write the whole wave function as
\begin{equation}
\psi_\pm ({\bf r}) = \psi (z) (1 \pm \alpha (p) \, p x) \; ,
\label{eq:totwf}
\end{equation}
where $\psi(z)$ is defined in Eq.~(\ref{eq:eikwfz}). The coefficient $\alpha (p)$ is 
largely energy dependent~\cite{bgprbook}. For sake of simplicity (and noting that the 
overall spin-orbit effect is not an increasing function of $p$ at large 
energies~\cite{bgprbook}), we rewrite the asymmetric factor as $\alpha (p) \, p \equiv 
\eta (p)$. Since $\eta R \ll 1$ at large energies, we may approximate 
$1 \pm \eta  x \approx \exp (\pm \eta x)$ and write the components of the projectile wave 
function as  
\begin{equation}
\psi_\pm ({\bf r}) \approx  e^{i p z - p_I z \pm \eta x} \; ,
\label{eq:totwf2}
\end{equation}
which correspond to two plane waves with complex momenta
${\bf P}_\pm = P\,\hat{z} \mp P_x \, \hat{x} = (p + i p_I) \, \hat{z} \mp i \eta \, 
\hat{x}$, respectively. 

We can now repeat the previous steps leading to Eq.~(\ref{eq:dwampl}) and write the
scattering amplitude as
\begin{equation}
a_\pm (p+i p_I, i \eta ) = \int d{\bf k}\, A((p+i p_I) \hat{z} \mp i \eta \hat{x},
{\bf k}) \; .
\label{eq:sodwampl}
\end{equation}
Again, following Eqs.~(\ref{eq:xsect}-\ref{eq:fsi}) we can perform a Taylor expansion 
including only terms linear in $p_I$ and $\eta$, namely
\begin{equation}
a_{\pm\,n}(P=p+i p_I, P_x=i\eta) \approx a_{\pm\,n}(p) + i p_I \, 
\frac{\partial a_{\pm\,n}}{\partial P} \Big\vert_{p_I=\eta=0} \mp i \eta \, 
\frac{\partial a_{\pm\,n}}{\partial P_x} \Big\vert_{p_I=\eta=0} \; ,
\label{eq:sotaylor}
\end{equation}
which leads to the response
\begin{equation}
W_\pm (p,p_I,\eta) = W_\pm^o (p) - 2 p_I \, {\rm Im} \left[ G_\pm^L (p) \right] \pm 
2 \eta \, {\rm Im} \left[ G_\pm^T (p) \right] \; ,
\label{eq:soxsect2}
\end{equation}
where
\begin{equation}
G_\pm^L (p) = \sum_n \, \left( a_{\pm\,n}^*\,\frac{\partial a_{\pm\,n}}{\partial P} 
\right) \Big\vert_{p_I=\eta =0} \quad , \quad 
G_\pm^T (p) = \sum_n \, \left( a_{\pm\,n}^*\, 
\frac{\partial a_{\pm\,n}}{\partial P_x} \right) \Big\vert_{p_I=\eta =0} \; .
\label{eq:sofsi}
\end{equation}

\section{Azimuthal asymmetry in unpolarized Drell-Yan}
\label{sec:asym}

When building an azimuthal asymmetry from a hard event, the response $W_\pm$ must
depend also on a variable, say $\xi$, which is characteristic of the asymmetric
distribution of the final products. In the example of Fig.~\ref{fig:dyframe}, $\xi$ can be
the angle $\phi$ which identifies the azimuthal direction of the lepton pair production
in the Collins-Soper frame. By making this dependence 
explicit in $W_\pm$, a typical feature of these asymmetric distributions for a free proton 
target (or when no distortion is taken into account, i.e. for $p_I = \eta = 0$, hence the
superscript $^o$ in the following formula) is that for each $\xi$ there exists a correlated 
$\xi'$ such that 
\begin{equation}
W_+^o (\xi ,p) = W_-^o (\xi',p) \; .
\label{eq:xiasym}
\end{equation}
In the previous example, the correspondence $\xi \rightarrow \xi'$ could be 
$\phi \rightarrow -\phi$ or $\phi \rightarrow \phi +\pi$. In other words, there are always 
two alternative ways for defining and/or calculating an azimuthal spin asymmetry when the 
effect of nuclear medium is not included: fix $\xi$ and reverse the spin of the 
projectile, or fix its spin and replace $\xi$ by $\xi'$.

Then, it is natural to define the spin asymmetry for a polarized Drell-Yan event on a free
proton target as 
\begin{equation}
A_{TT} = \frac{W_+^o(\xi ,p) - W_-^o (\xi ,p)}{W_+^o(\xi ,p) + W_-^o (\xi ,p)}
= \frac{W_+^o (\xi ,p) - W_+^o (\xi' ,p)}{W_+^o (\xi ,p) + W_+^o (\xi' ,p)} \; .
\label{eq:ssa}
\end{equation}
In addition, we also define the "polarization averaged" asymmetry
\begin{equation}
\bar{A}^o = \frac{1}{2} \, \Bigg( W_+^o(\xi ,p) + W_-^o(\xi ,p) - 
W_+^o(\xi',p) - W_-^o(\xi',p) \Bigg) \; .
\label{eq:uxiasym}
\end{equation}
Because of Eq.~(\ref{eq:xiasym}), the latter $\bar{A}^o$ turns out to vanish. It means 
that those scattering amplitudes that contribute to $A_{TT}$ simply average to zero. 
Effects of a different origin may create a nonvanishing $\bar{A}^o$. In fact, because of 
the nuclear damping summarized in Eq.~(\ref{eq:soxsect2}), the analogue of 
Eq.~(\ref{eq:xiasym}) for a real nuclear target becomes
\begin{equation}
W_+ (\xi, p,p_I,\eta) - W_- (\xi', p, p_I,\eta) = 2p_I \, {\rm Im} \left[ G_-^L
(\xi', p) - G_+^L (\xi, p) \right] + 2\eta \, {\rm Im} \left[ G_+^T (\xi, p) + 
G_-^T (\xi', p) \right] \; ,
\label{eq:soxiasym}
\end{equation}
where the "undistorted" contributions $W_\pm^o$ cancel just because of
Eq.~(\ref{eq:xiasym}). If $p_I$ and/or $\eta$ are not vanishing, then also
Eq.~(\ref{eq:soxiasym}) is not vanishing. Consequently, the generalization of $\bar{A}^o$ 
to the case of a real nuclear target reads
\begin{eqnarray}
\bar{A} &= &\frac{1}{2} \, \Bigg( W_+(\xi ,p,p_I,\eta) + W_-(\xi ,p,p_I,\eta) - 
W_+(\xi',p,p_I,\eta) - W_-(\xi',p,p_I,\eta) \Bigg) \nonumber \\
&= &\frac{1}{2} \, \Bigg\{ 2p_I \, {\rm Im} \left[ G_+^L (\xi', p) + 
G_-^L (\xi', p) - G_+^L (\xi, p) - G_-^L (\xi, p) \right] 
\nonumber \\
& &\quad + 2\eta \, {\rm Im} \left[ G_+^T (\xi, p) - G_-^T (\xi, p) - 
G_+^T (\xi', p) + G_-^T (\xi', p) \right] \Bigg\} \; .
\label{eq:souxiasym}
\end{eqnarray}

When nuclear distortion is neglected or a free proton target is considered, i.e. for
$p_I = \eta = 0$, then Eq.~(\ref{eq:xiasym}) holds and $\bar{A} \equiv \bar{A}^o = 0$. It 
means that for a nuclear target the projectile can equivalently follow different paths 
($\xi$ and $\xi'$) inside the nucleus. When including the nuclear distortion, from 
Eq.~(\ref{eq:souxiasym}) we get in general that $\bar{A} \neq 0$: because of the nuclear 
damping, the two paths $\xi$ and $\xi'$ are not equivalent and the nucleus acts as a 
polarizer for the projectile. 

We can now put this picture in the physical context described in Sec.~\ref{sec:intro}. If
the Lam-Tung sum rule were correct, the cross section would not depend on the azimuthal
$\phi$ distribution of the final lepton pair in an unpolarized Drell-Yan event in nuclear
targets. This is somewhat the same content of Eq.~(\ref{eq:xiasym}) and of the condition 
$\bar{A}^o = 0$ for $\xi = \phi$. However, as already stressed in Sec.~\ref{sec:intro} the 
Lam-Tung sum rule is valid only for the direction of annihilation along the $\hat{z}$ axis 
(see Fig.~\ref{fig:dyframe}), i.e. for collinear partons in the hard event. Violation of 
the sum rule emphasizes the role of transverse dynamics of partons inside hadrons. A 
similar azimuthal asymmetry can be obtained by considering that the transverse momentum of
the Drell-Yan pair originates from the transverse momentum of the colliding hadron when it
is deflected by nuclear damping (particularly by the "transversely asymmetric" spin-orbit 
effect). This sort of duality that is intriguingly established between the partonic and 
hadronic interpretations of azimuthal asymmetries in terms of transverse (spin) degrees of 
freedom, might suggest that spin-orbit effects play a role also at the elementary level in
characterizing the transverse dynamics of partons inside hadrons.

In conclusion, when a Drell-Yan hard event is considered in a collision of a hadronic
unpolarized projectile with an unpolarized nuclear target, the nuclear damping and the 
soft spin-orbit interaction happening on the nuclear surface can distort the projectile 
wave function and produce an azimuthal asymmetric distribution of the final lepton pair 
$(\bar{A} \neq 0)$ even if there is no asymmetry in the corresponding collision on a 
free proton target $(\bar{A}^o = 0)$. The necessary condition for this to happen is that 
for the corresponding Drell-Yan with a polarized projectile and/or proton target the 
transverse spin asymmetry is not vanishing $(A_{TT} \neq 0)$, because both observables are
expressed in terms of the same scattering amplitudes $a_n$.

\section{Discussion and perspectives}
\label{sec:size}

Given the results summarized in the last paragraph of previous section, the question 
arises about the interpretation of the experimental data of 
Refs.~\cite{Falciano:1986wk,Guanziroli:1987rp,Conway:1989fs,Anassontzis:1987hk} and of 
possible future Drell-Yan experiments with (anti)proton beams at the foreseen HESR at 
GSI: when a nuclear target is employed, how big is the distortion of the beam wave 
function at a given kinematics? And how can this effect be disentangled from the genuine 
azimuthal asymmetry due to the hard event?

Concerning the first question, the main problem is the lack of knowledge about the
microscopic mechanisms that build up the elementary scattering amplitudes $a_n$ in
Eq.~(\ref{eq:sofsi}). We can attempt some educated guess based on simplifying hypothesis 
and on the general properties of known parton distributions and of hadron-hadron 
phenomenology. 

For sake of simplicity, let $W^o$ in Eq.~(\ref{eq:xsect}) be dominated by a single 
amplitude $a(P,P_x) = a_{_R}+ i a_{_I}$ with $\partial a_{_I} / \partial P_{(x)} \approx 
0$ and $a_{_R} \approx 0$ at $p_I=\eta=0$. Then,
\begin{equation}
W^o(p) \approx a_{_I}^2 \quad , \quad G^i(p) \approx \left( a_{_I}\, 
\frac{\partial a_{_R}}{\partial P_i} \right) \Big\vert_{p_I=\eta =0}\; ,
\label{eq:1ampl}
\end{equation}
where $i=L,T$ correspond to longitudinal and transverse components with $P_L\equiv P = p+ i
p_I$ and $P_T \equiv P_x = i \eta$. 

We define the ratios
\begin{equation}
r_I = \frac{2 \, p_I \, G^L(p)}{W^o(p)}\quad , \quad r_{so} = \frac{2\, \eta \,
G^T(p)}{W^o(p)} \; ,
\label{eq:ratios}
\end{equation}
that should tell us about the relative size between the asymmetry of final lepton pairs
generated by the collision of an unpolarized hadron beam on a nucleus, and by the 
collision of the same beam, but polarized, on a free proton target. In other words, $r_I$ 
and $r_{so}$ should tell us how big is the asymmetry produced by nuclear damping and 
spin-orbit, respectively, with respect to the asymmetry produced by the genuine polarized 
Drell-Yan hard event. 

In order to estimate $r_I$ and $r_{so}$, we observe that the parton densities significantly
change over the scale of their parent hadron size $R_h \sim 0.2 \div 0.5$ fm (which
corresponds to a momentum scale of $p_h \sim 1 \div 0.4$ GeV/$c$, respectively). We assume 
that also the amplitude $a$ changes significantly only over the same scale; neglecting 
fine details and looking only to the order of magnitudes, we can roughly conclude that 
\begin{equation}
\frac{G^i(p)}{W^o(p)} \approx \left( \frac{1}{a_{_I}}\, \frac{\partial a_{_R}}{\partial 
P_i} \right) \Bigg\vert_{p_I=\eta=0} \sim \frac{1}{p_h} \sim R_h \; .
\label{eq:stimaratio}
\end{equation}
In other words, we safely assume that the relevant mechanisms involve correlations at most 
over the scale $R_h$ excluding long-range interactions that would imply large derivatives 
of $a$ at $p_I = \eta = 0$ and, consequently, large values of $G^i$. In this hypothesis, 
the ratios of Eq.~(\ref{eq:ratios}) become $r_I \sim 2 p_I / p_h$ and $r_{so}\sim 2 \eta / 
p_h$, respectively. As already mentioned in Sec.~\ref{sec:eik}, experimental data about 
nuclear transparency on $^{12}$C seem to suggest $p_I \sim 50$ MeV/$c$. This value can be 
interpreted as the mean free path of a hadron travelling in the nuclear medium, and it 
should not depend on the mass number but for very light nuclei. The corresponding ratio 
$r_I$, then, could be in the range $5 \div 10$ \%. 

In order to estimate $\eta$, it is useful to consider the nuclear analyzing power $A_N$, 
i.e. the asymmetry between cross sections at a given transferred momentum for spin up and 
spin down projectiles colliding on an unpolarized nucleus. If the typical range associated 
with spin-orbit effects is given by the surface thickness, we can approximate this scale in 
a nucleus again with $R_h$. Hence, we can assume that $\langle A_N \rangle \approx \eta \, 
R_h \sim \eta / p_h$, where $\langle A_N \rangle$ is the analyzing power averaged over the 
transferred momenta relevant to the ISI occurring during the propagation of the projectile 
inside the nuclear target, typically in the range $0 \div 0.5$ GeV/$c$. The problem of 
estimating $r_{so}$ is, therefore, translated into the determination of the typical size of
$\langle A_N \rangle$. In the literature, recent measurements and attempts of theoretical
interpretations have been published for elastic scatterings at energies beyond 10 GeV 
with proton-proton~\cite{Akchurin:1989er,Buttimore:1998rj}, 
proton-$^{12}$C~\cite{Tojo:2002wb,Kopeliovich:2000kz,Selyugin:2003xc}, and 
antiproton-proton~\cite{Akchurin:1989er} systems (for a collection of older experimental
data, see Ref.~\cite{Buttimore:1998rj} and references therein). Results for $A_N$ with 
proton and $^{12}$C targets are similar in magnitude and shape and they show a steep 
decrease from 10\% at 10 GeV down to $2\div 4$\% at 20 GeV, with somewhat higher values at 
larger energies around 30 GeV but with large error bars. At large energies, the 
antiproton-proton results have also a similar magnitude to the previous 
ones~\cite{Akchurin:1989er}. However, extrapolation of these findings to other nuclear 
targets (maybe, isoscalar nuclei like $^{12}$C) should be taken with great care, since the
reproduction of "nuclear" $A_N$ in terms of proton-proton and antiproton-proton amplitudes
is not theoretically well established~\cite{Kopeliovich:2000kz,Selyugin:2003xc}. For 
transferred momenta in the range of interest $0\div 0.5$ GeV/$c$, $A_N$ already gets its 
maximum value for almost all beam energies and the other values are not much smaller than 
that. At larger angles, $A_N$ can reach large values at all energies, but this 
situation is not relevant here and it won't be considered. In conclusion, with the above 
caveat we can assume $\langle A_N \rangle \approx A_N$ and finally deduce that the 
approximate size of the spin-orbit nuclear damping is in the range $4\div 10$\%, depending 
on the projectile energy. 

The previous estimates about $r_I$ and $r_{so}$ are rather conservative, since they are 
based on hard interactions occurring at most within the short range $R_h$. However, we 
cannot exclude the case of a scattering amplitude $a$ having a much steeper dependence on 
small values of $P (P_x)$ because of correlations at a longer range. For example, if $a$ 
has a resonance-like behaviour with a pole as 
\begin{equation}
a(k) = \frac{\alpha}{k - i \gamma} \; ,
\label{eq:pole}
\end{equation}
where $k=P$ or $P_x$, then from Eq.~(\ref{eq:1ampl}) we get
\begin{equation}
r_I = \frac{2 \, p_I \, G^L(p)}{W^o(p)} = \frac{2 \, p_I}{\gamma} \quad , \quad r_{so} = 
\frac{2\, \eta \, G^T(p)}{W^o(p)} = \frac{2\,\eta}{\gamma} \; .
\label{eq:ratiospole}
\end{equation}
In the "short-range" hypothesis, $\gamma \sim 1/R_h$. But if $a$ represents an intermediate 
state with a narrow width, then the nuclear damping is largely enhanced. This is not 
surprising, since a narrow-width resonance would propagate inside the nucleus with a mean 
free path $\sim 1/\gamma$ increasing the importance of nuclear effects. A possible example 
of this phenomenon is represented by the Drell-Yan process at the $J/\psi$ 
mass~\cite{Anselmino:2004ki}, but involving a nuclear target. Since this resonance is very 
narrow and it is present in all possible amplitudes, if the process is studied via a 
hadron-nucleus collision we would have the standard conditions of nuclear shadowing, i.e. 
the propagation of a quasi-real particle through large distances inside the nucleus.

In summary, nuclear damping effects in hadron-nucleus collisions will affect azimuthal 
asymmetries from Drell-Yan hard events most likely by few percents, unless quasi-real 
intermediate states are involved in the propagation of the hadron projectile inside the 
target nucleus before the hard event. This condition is not so unlikely, if higher-twist 
contributions are relevant in Drell-Yan processes. Indeed, if $M_p/M$ is related to the 
ratio between subleading and leading twists (with $M_p$ the proton mass), a simulation
in the typical kinematics foreseen at HESR at GSI~\cite{Bianconi:2004??} reveals that the 
cross section has a steep $M$ dependence with most events concentrated at the lowest 
allowed $M = 4$ GeV, where $M_p/M \approx 25$\%. However, new measurements of Drell-Yan 
events are of course needed to confirm and substantiate previous arguments, since the 
available data pertain a completely different 
regime~\cite{Falciano:1986wk,Guanziroli:1987rp,Conway:1989fs,Anassontzis:1987hk}, where, 
incidentally, large error bars in the azimuthal asymmetries do not exclude nuclear effects 
when changing from the deuteron to the tungsten targets (see Fig.8 of 
Ref.~\cite{Guanziroli:1987rp}). 

Going back to the beginning of the section, we consider the second question about the 
possibility of disentangling the effect of nuclear damping from the genuine azimuthal 
asymmetry produced by a Drell-Yan hard event. Since spin-orbit effects are mostly 
associated with a single quasi-elastic projectile-nucleon scattering, detection of the 
Drell-Yan lepton pair in coincidence with a recoil proton with small longitudinal momentum 
($< 300$ MeV/$c$) and large transverse momentum ($400 \div 800$ MeV/$c$) would allow for a 
quantitative check of the importance of ISI (responsible for this large-angle emission) 
and for an indirect determination of the polarization of the projectile before the hard 
event, because this would determine the direction of the recoil proton momentum. With this
tagging technique, it should be possible, in principle, to study a polarized Drell-Yan 
event from unpolarized beams and targets.


\section{Conclusions}
\label{sec:end}

Since several years, a series of measurements of Drell-Yan events in unpolarized high-energy
hadronic collisions on nuclei have been published in the 
literature~\cite{Falciano:1986wk,Guanziroli:1987rp,Conway:1989fs,Anassontzis:1987hk} that 
still await for a coherent quantitative explanation. In fact, the differential cross section
shows an unexpected largely asymmetric azimuthal distribution of the final lepton pair with
respect to the production plane (see Fig.~\ref{fig:dyframe}). At present, the most promising
interpretation identifies the source of the observed $\cos 2\phi$ asymmetry in a
leading-twist contribution to the cross section containing a peculiar distribution function,
that describes the intrinsic transverse momentum distribution of transversely polarized
partons inside unpolarized hadrons~\cite{Boer:1999mm}. 

In this paper, we have explored the consequences of the fact that the above experiments
involve target nuclei, namely that the hadronic beam (pions or antiprotons) has to cross
the nuclear surface and travel inside the nuclear matter for some length (related to its
energy) before colliding on a bound nucleon to produce the Drell-Yan hard event. The wave
function of the projectile is distorted by the nuclear damping and by spin-orbit
interactions happening on the nuclear surface, producing also an azimuthal asymmetry in the
distribution of the Drell-Yan lepton pair; this effect should be added to the one
originating from the elementary hard event. The necessary condition for this to happen is
that for the corresponding Drell-Yan with polarized projectile and/or free proton target the
asymmetry produced by flipping one of the spins is not vanishing, since both asymmetries are
expressed in terms of the same elementary scattering amplitudes. 

Using some educated guess based on simplifying hypothesis and on the general properties of
known parton distributions and of hadron-hadron phenomenology, we estimated that the nuclear
damping should affect the azimuthal asymmetry from Drell-Yan hard events most likely by few
percents, unless quasi-real intermediate states are involved in the propagation of the 
hadron projectile inside the target nucleus before the hard event. This condition is not 
so unlikely if nuclear targets will be possibly adopted at the HESR ring at GSI, where in
the considered kinematics either higher-twist contributions are relevant or the invariant 
mass of hadronic resonances with narrow width is reached in the annihilation (see 
Refs.~\cite{panda,pax,pax2,assia,Efremov:2004qs,Anselmino:2004ki,Radici:2004ij} for 
further details). 

It should be possible to isolate "nuclear" asymmetries from "hard" asymmetries by
considering a Drell-Yan process where the final lepton pair is detected in coincidence with
a recoil proton with small longitudinal momentum and large transverse momentum. In fact,
only the rescatterings of the hadron beam propagating inside the nuclear medium would be
responsible of such large-angle emission. It is interesting to note that, as the $\cos
2\phi$ asymmetry could be produced by the transverse dynamics of partons inside hadrons
leading to a violation of the Lam-Tung sum rule~\cite{Lam:1980uc}, similarly the same
asymmetry can be obtained by considering the transverse momentum of the colliding hadron 
when it is deflected by nuclear damping (particularly by the "transversely asymmetric" 
spin-orbit effect). This parallelism might suggest that spin-orbit effects play a role 
also at the elementary level in characterizing the transverse dynamics of partons inside 
hadrons.


\bibliographystyle{apsrev}
\bibliography{eik,ssa}


\end{document}